\documentclass[11pt,a4paper]{article}

\usepackage[dvips]{graphicx}

\usepackage[cp1251]{inputenc}
\usepackage[T2A]{fontenc}
\usepackage[english,russian]{babel}

\usepackage{animate}

\usepackage{amsmath}
\usepackage{amssymb}
\usepackage{amsxtra}
\usepackage{amsthm}
\usepackage{latexsym}
\usepackage{array}
\usepackage{multirow}
\usepackage{hhline}
\usepackage{color}
\usepackage{longtable}
\numberwithin{equation}{section}
\theoremstyle{plain}

\theoremstyle{remark}

\newcommand{\rank}{\mathop{\rm rank}\nolimits}
\usepackage{hyperref}
\begin{document}

\title{On one unstable bifurcation in the dynamics of vortex structure}

\author{P.\,E.~Ryabov}

\date{}

\maketitle

\begin{abstract}
In this paper we consider a completely Liouville integrable Hamil\-to\-nian system with two degrees of freedom, which describes the dynamics of two vortex filaments in a Bose-Einstein condensate enclosed in a cylindrical trap. For vortex pairs of positive intensity detected bifurca\-tion of three Liouville tori  into one. Such bifurcation was found in the integrable case of Goryachev-Chaplygin-Sretensky in the dynamics of a rigid body. For the integrable perturbation of the physical parameter of the intensity ratio,
identified bifurcation proved to be unstable, which led to bifurcations of the type of two tori into one and vice versa.

\vspace{3mm}


\end{abstract}
{

\parindent=0mm

УДК 532.5.031, 517.938.5


------------------------------------------------------------

Received on 10 December 2018.

------------------------------------------------------------

The work was supported by RFBR
grants 16-01-00170 and  17-01-00846.

------------------------------------------------------------

Ryabov Pavel Evgen`evich

PERyabov@fa.ru

Financial University under the Government of the Russian Federation\\
Leningradsky prosp. 49, Moscow, 125993 Russia

Institute of Machines Science, Russian Academy of Sciences\\
Maly Kharitonyevsky per. 4, Moscow, 101990 Russia

Udmurt State University\\
ul. Universitetskaya 1, Izhevsk, 426034 Russia

---------------------------------------------------------------
}

\tableofcontents

\section{Введение}
В аналитической динамике вихревых структур особое место уделяется интегрируемым моделям точечных вихрей на плоскости. С другой стороны,
исследования динамики вихрей, но уже в квантовой физике, показали, что квантовые вихри ведут себя примерно также, как тонкие вихревые нити, изучаемые в классической гидродинамике. Особое место при этом занимают вихревые структуры в бозе-эйнштейновском конденсате,  полученном для ультрахолодных атомных газов. Cовременный обзор таких исследований представлен, например, в \cite{fett2009}. В данной статье будет рассмотрена математическая модель динамики двух вихревых нитей в бо\-зе-эйн\-штей\-нов\-ском кон\-ден\-са\-те, заключенном в цилиндрической ловушке \cite{kevrekPhysLett2011}.  Такая модель приводит к вполне интегрируемой по Лиувиллю гамильтоновой системе с двумя степенями свободы, и по этой причине, могут быть применены  топологические методы, используемые в таких задачах. Топологические методы  успешно показали себя в исследовании характера устойчивости абсолютных и относительных хореографий \cite{borkil2000}, \cite{bormamkil2004}, \cite{kilinbormam2013}, \cite{BorSokRyab2016}. Таким движениям в интегрируемых моделях, как правило,  соответствуют  значения постоянных первых интегралов, для которых сами интегралы, рассматриваемые как функции от фазовых переменных, оказываются зависимыми в смысле линейной зависимости дифференциалов. Основную роль в топологическом анализе играет бифуркационная диаграмма отображения момента. В работах \cite{SokRyabRCD2017, sokryab2018} получена бифуркационная диаграмма  в задаче о движении системы двух точечных вихрей в бозе-эйнштейновском конденсате и исследованы бифуркации лиувиллевых торов, в частности, наблюдалась бифуркация двух торов в один.

Настоящая публикация посвящена интегрируемому возмущению динамики рассматриваемой модели. В данной работе исследованы бифуркации торов Лиувилля, в частности, в случае вихревой пары положительной  интенсивности для некоторых значений физических параметров наблюдается бифуркация трех торов в один. Бифуркации  трех торов в один отмечена в работах М.\,П.~Харламова для интегрируемого случая Чаплыгина -- Горячева -- Сретенского в динамике твердого тела (см.~\cite{Kharlamov1988})  и как одна из особенностей в виде так называемого 2-атома особого слоя лиувиллева слоения в работах А.\,Т.~Фоменко, А.\,В.~Болсинова, C.\,В.~Матвеева \cite{bolsmatvfom1990}. В работе А.\,А.~Ошемкова и М.\,А.~Тужилина \cite{oshtuzh2018}, посвященной расщепляемости седловых особенностей, такая бифуркация оказалась неустойчивой и приведены ее возмущенные слоения, одна из которых и реализуется в рассматриваемом случае.

\section{Модель и необходимые определения}
Математическая модель динамики двух точечных вихрей в Бо\-зе-Эйн\-штей\-нов\-ском кон\-ден\-са\-те, заключенном в цилиндрической ловушке, описывается системой дифференциальных уравнений, которая может быть представлена в гамильтоновой форме \cite{kevrekPhysLett2011}:
\begin{equation}
\label{x1}
\boldsymbol{\dot{\zeta}}=\{\boldsymbol\zeta, H\}
\end{equation}
c функцией Гамильтона
\begin{equation}
\label{x2}
H=\ln[1-(x_1^2+y_1^2)]+a^2\ln[1-(x_2^2+y_2^2)]-a b\ln[(x_2-x_1)^2+(y_2-y_1)^2].
\end{equation}
Здесь через $(x_k,y_k)$ обозначены декартовы координаты $k$-ого вихря ($k=1,2$), фазовый вектор $\boldsymbol\zeta$ имеет координаты $\{x_1,y_1,x_2,y_2\}$, параметр $a$ обозначает отношение интенсивностей $\frac{\Gamma_2}{\Gamma_1}$. Физический параметр $b$ характеризует меру вихревого взаимодействия интенсивностей и пространственной неоднородности в гармонической  ловушке \cite{kevrekPhysLett2011}, \cite{kevrek2013}. В работах \cite{SokRyabRCD2017}, \cite{sokryab2018} такой параметр принимался равным единице, тем не менее в ряде физических работ \cite{kevrekPhysLett2011}, \cite{kevrek2013} для параметра $b$ в случае вихревой пары положительной интенсивнойсти ($a=1$) на основе экспериментальных данных принимались следующие значения $b=2$, $b=1,35$, $b=0,1$  и др. В связи с чем представляет интерес исследование фазовой топологий, когда параметр взаимодействия $b$ принимает любые положительные значения.

Фазовое пространство $\cal P$ задается в виде прямого произведения двух открытых кругов радиуса $1$, с выколотом множеством столкновениий вихрей
\begin{equation*}
\label{x3}
{\cal P}=\{(x_1,y_1,x_2,y_2)\,:\, x_1^2+y_1^2<1, x_2^2+y_2^2<1\}\smallsetminus \{x_1=x_2, y_1=y_2\}.
\end{equation*}
Пуассонова  структура на фазовом пространстве $\cal P$ задается в стандартном виде
\begin{equation}
\label{x4}
\{x_i,y_j\}=\frac{1}{\Gamma_i}\delta_{ij},
\end{equation}
где $\delta_{ij}$ -- символ Кронекера.

Система \eqref{x1} допускает один дополнительный первый интеграл движения -- \textit{момент завихренности}
\begin{equation*}
\label{x5}
F=x_1^2+y_1^2+a(x_2^2+y_2^2).
\end{equation*}

Функция $F$ вместе с гамильтонианом $H$ образуют на $\cal P$ полный инволютивный набор интегралов системы $\eqref{x1}$.
Согласно теореме Лиувилля-Арнольда  регулярная поверхность уровня первых интегралов вполне интегрируемой гамильтоновой системы представляет собой несвязное объединение торов, заполненных условно-периодическими траекториями. Определим \textit{интегральное отображение} ${\cal F}\,:\, {\cal P}\to {\mathbb R}^2$, полагая $(f,h)={\cal F}(\boldsymbol\zeta)=(F(\boldsymbol\zeta), H(\boldsymbol\zeta))$. Отображение $\cal F$ принято также называть \textit{отображением момента}. Обозначим через $\cal C$ совокупность всех критических точек отображений момента, то есть
точек, в которых $\rank d{\cal F}(x) < 2$. Множество критических значений $\Sigma = {\cal F}({\cal C}\cap{\cal P})$ называется \textit{бифуркационной диаграммой}.

\section{Бифуркационная диаграмма}
Бифуркационная диаграмма $\Sigma$ отображения момента $\cal F$ была явно определена в работе \cite{ryab_arxiv2018}. В случае неравных положительных интенсивностей, т.е. когда парамер отношения интенсивностей $a$ отличен от единицы, бифуркационная диаграмма $\Sigma_{a,b}$ имеет вид:
\begin{equation}
\label{x2_4}
\Sigma_{a,b}:\left\{
\begin{array}{l}
f=(1+at^2)r_1^2,\\[3mm]
h=\ln(1-r_1^2)+a^2\ln(1-t^2r_1^2)-ab\ln[(1+t)^2r_1^2],\\[3mm]
\displaystyle{r_1^2=\frac{abt^3+(a-b-1)t^2+(a-1+ab)t-b\pm\sqrt{\cal D}(t+1)}{2t[-t^3+(ab-1)t^2+(a-b)t+a]}.}
\end{array}\right.
\end{equation}

В случае вихревой пары положительной интенсивности, т.е. когда параметр отношения интенсивностей $a$ равен единице, бифуркационная диаграмма $\Sigma_{1,b}$ принимает простой вид и состоит из двух кривых $\gamma_1$ и $\gamma_2$, где
\begin{equation}\label{x2_5}
\begin{array}{l}
\displaystyle{\gamma_1: h=2\ln\left(1-\frac{f}{2}\right)-b\ln(2f),\quad 0<f<2;}\\
\gamma_2: \left\{
\begin{array}{l}
\displaystyle{h=\ln\left[\frac{s^2(s-1)}{b+s-1}\right]-b\ln\left[\frac{bs^2}{b+s-1}\right],}\\[3mm]
\displaystyle{f=\frac{bs^2-2(s-1)(b+s-1)}{b+s-1},}
\end{array}\right.\qquad s\in \left(1;\frac{2(1+\sqrt{b})}{2+\sqrt{b}}\right].
\end{array}
\end{equation}
Отметим, что при значениях физического параметра $b>3$, кривая $\gamma_2$ имеет точку возврата при $s_{cusp}=\frac{\left[2-b+\sqrt{b(b-2)}\right](b-1)}{b-2}$, которая совпадает с точкой касания для $b=3$ при $s_{touch}=\frac{2(1+\sqrt{b})}{2+\sqrt{b}}$.

Параметризованная кривая \eqref{x2_4} также имеет точки возврата, которые удовлетворяют уравнению $P(t)=0$. Здесь $P(t)$ -- многочлен десятой степени, коэффициенты  которого зависят от физических параметров $a$ и $b$. Более того, дискриминант этого многочлена описывает ситуацию, когда точки возврата ``сливаются'' в одну и одна из ветвей становится гладкой. Такая ситуация наблюдается в динамике взаимодействия двух точечных вихрей в идеальной жидкости внутри цилиндра \cite{BorMamSokolovskii2003}. Далее, вновь происходит рождение ``точек возврата''.  На рис.~1 и 2 представлены при фиксированном значении физического параметра $b>3$ увеличенный фрагмент бифуркационной диаграммы, которая соответствует значению физического параметра $a=1$ и анимация бифуркационной диаграммы интегрируемого возмущения для $a\in [1;1,24]$. Знак $``+``$ на рис.~1 отвечает эллиптическим (устойчивым) периодическим решениям в фазовом пространстве, которые являются пределом концентрического семейства двумерных регулярных торов, а знак $``-``$  -- гиперболическим периодическим решениям, для которых существуют движения, асимптотические к этому решению, лежащие на двумерных сепаратрисных поверхностях. Как и следовало ожидать, смена типа происходит в точке касания $B$ и точке возврата $A$ бифуркационной диаграммы $\Sigma$.

\begin{figure}[!ht]
\centering
\includegraphics[width=1\textwidth]{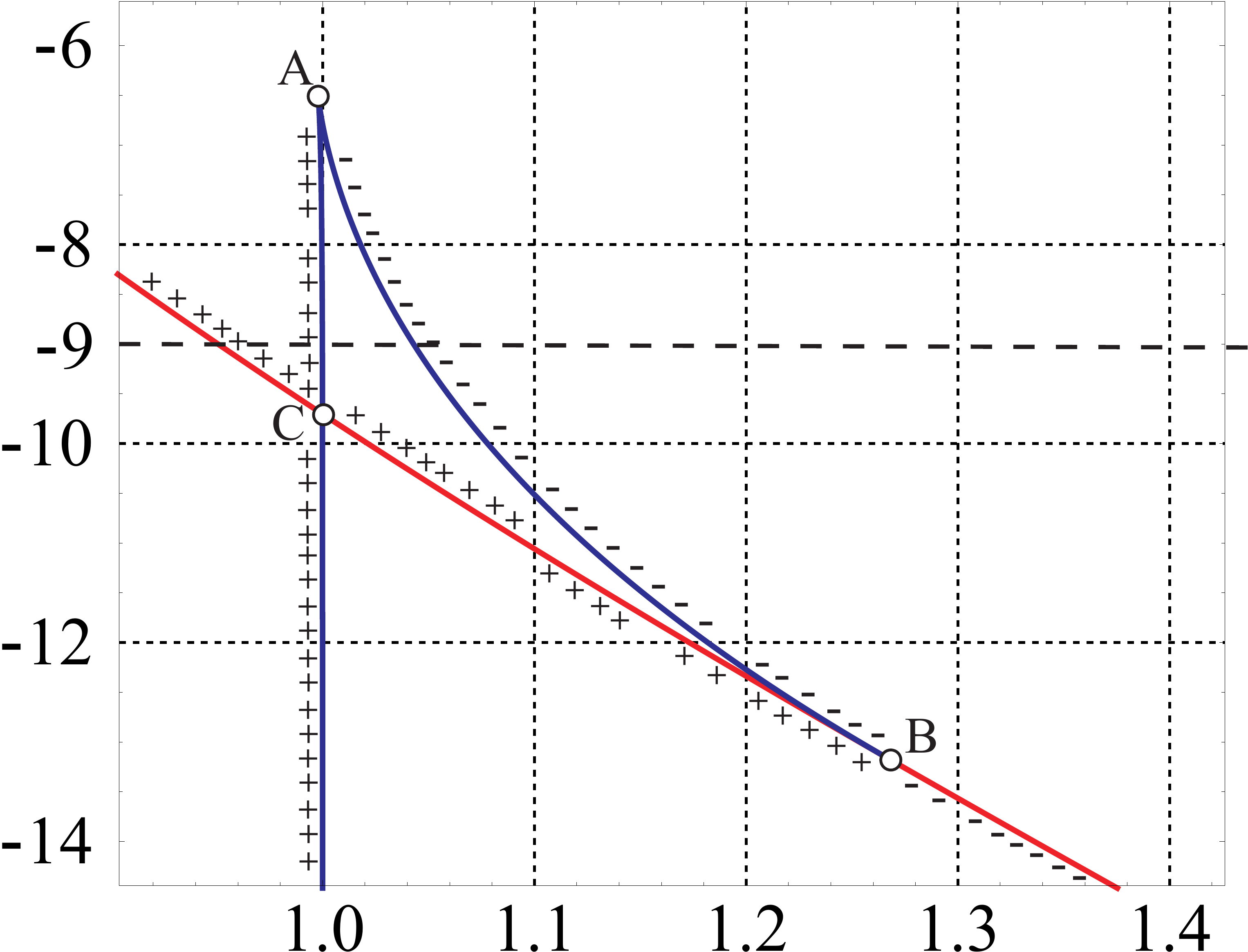}
\caption{Увеличенный фрагмент бифуркационной диаграммы $\Sigma$ для $a=1$ и $b>3$.}
\label{fig1}
\end{figure}

\begin{figure}[!ht]
\begin{center}
\animategraphics[width=12cm,height=10cm,controls,buttonsize=1em,buttonfg=0.5]{3}{D1_anim_}{001}{60}
\end{center}
\caption{Анимация увеличенного фрагмента бифуркационной диаграммы $\Sigma$ для $a\in [1;1,24]$ при фиксированном $b>3$.}
\label{fig2}
\end{figure}

\section{Бифуркация $3\mathbb T^2 \to \mathbb S^1\times\left(\mathbb S^1\,\dot{\cup}\,\mathbb S^1\,\dot{\cup}\,\mathbb S^1\right)\to \mathbb T^2$}
Здесь мы ограничимся положительными интенсивностями. Выполним явное приведение к системе с одной степенью свободы. Для этого в системе \eqref{x1} с гамильтонианом  \eqref{x2} перейдем от фазовых переменных $(x_k,y_k)$ к новым переменным $(u,v,\alpha)$ по формулам:
\begin{equation*}\label{z1}
\begin{array}{l}
x_1=u\cos(\alpha)-v\sin(\alpha),\quad y_1=u\sin(\alpha)+v\cos(\alpha),\\[3mm]
\displaystyle{x_2=\frac{1}{\sqrt{a}}\sqrt{f-u^2-v^2}\cos(\alpha),\quad y_2=\frac{1}{\sqrt{a}}\sqrt{f-u^2-v^2}\sin(\alpha).}
\end{array}
\end{equation*}

Физические переменные $(u,v)$ представляют собой декартовы координаты одного из вихрей в системе координат, связанной с другим вихрем, вращающейся вокруг центра завихренности. Выбор таких переменных подсказан наличием интеграла момента завихренности $F=\Gamma_1(x_1^2+y_1^2)+\Gamma_2(x_2^2+y_2^2)$, который инвариантен относительно группы вращений $SO(2)$. Существование однопараметрической группы симметрии позволяет выполнить редукцию к системе с одной степенью свободы, подобно тому, как это делается в механических системах с симметрией  \cite{Kharlamov1988}.  Обратная замена (в общем случае)
\begin{equation*}
U=\Gamma_1\frac{x_1x_2+y_1y_2}{\sqrt{x_2^2+y_2^2}},\quad V=\Gamma_1\frac{y_1x_2-x_1y_2}{\sqrt{x_2^2+y_2^2}}
\end{equation*}
приводит к каноническим переменным относительно скобки \eqref{x4}:
\begin{equation*}
\{U,V\}=-\{V,U\}=1,\quad \{U,U\}=\{V,V\}=0.
\end{equation*}

Система по  отношению к новым переменным $(u,v)$ является гамильтоновой
\begin{equation}\label{z2}
\dot u=\frac{\partial H_1}{\partial v},\quad
\dot v=-\frac{\partial H_1}{\partial u}
\end{equation}
с гамильтонианом
\begin{equation}
\label{z3}
\begin{array}{l}
\displaystyle{H_1=\ln(1-u^2-v^2)+a^2\ln\Bigl(1-\frac{f}{a}+\frac{u^2+v^2}{a}\Bigr)-}\\[3mm]
\displaystyle{-ab\ln\Bigl[\frac{f}{a}+\frac{a-1}{a}(u^2+v^2)-\frac{2}{\sqrt{a}}\sqrt{f-u^2-v^2}\,u\Bigr].}
\end{array}
\end{equation}

Угол поворота $\alpha(t)$ вращающейся системы координат удовлетворяет дифференциальному уравнению
\begin{equation*}
\displaystyle{\dot\alpha=\frac{R(u,v)}{Q(u,v)},}
\end{equation*}
где
\begin{equation*}
\begin{array}{l}
R(u,v)=-2a\bigl\{\sqrt{f-u^2-v^2}[(a^2-a+b)(u^2+v^2)+(a-b)f+ab]-\\[3mm]
-\sqrt{a}u[(b-2a)(u^2+v^2)+(2a-b)f+ab]\bigr\},\\[3mm]
Q(u,v)=\sqrt{f-u^2-v^2}\bigl\{2\sqrt{a}u\sqrt{f-u^2-v^2}(u^2+v^2-f+a)-\\[3mm]
-(u^2+v^2-f+a)[(a-1)(u^2+v^2)+f]\bigr\}.
\end{array}
\end{equation*}

Неподвижные точки редуцированной системы \eqref{z2} определяются критическими точками приведенного гамильтониана \eqref{z3} и соответствуют  относительным равновесиям вихрей в системе \eqref{x1}. Для фиксированного значения интеграла момента завихренности $f$ регулярные уровни приведенного гамильтониана \eqref{z3} -- компактны и движения происходят по замкнутым кривым.
Можно показать, что критические значения приведенного гамильтониана определяют бифуркационную диаграмму \eqref{x2_4} и \eqref{x2_5}. Для участка бифуркационной кривой $(AB)$ движение на плоскости $(u,v)$ происходит по кривой, которая топологически устроена как $\mathbb S^1\,\dot{\cup}\,\mathbb S^1\,\dot{\cup}\,\mathbb S^1$ (рис.~3), а интегральная критическая поверхность представляет собой тривиальое расслоение над $\mathbb S^1$ со слоем $\mathbb S^1\,\dot{\cup}\,\mathbb S^1\,\dot{\cup}\,\mathbb S^1$.

\begin{figure}[!ht]
\centering
\includegraphics[width=0.7\textwidth]{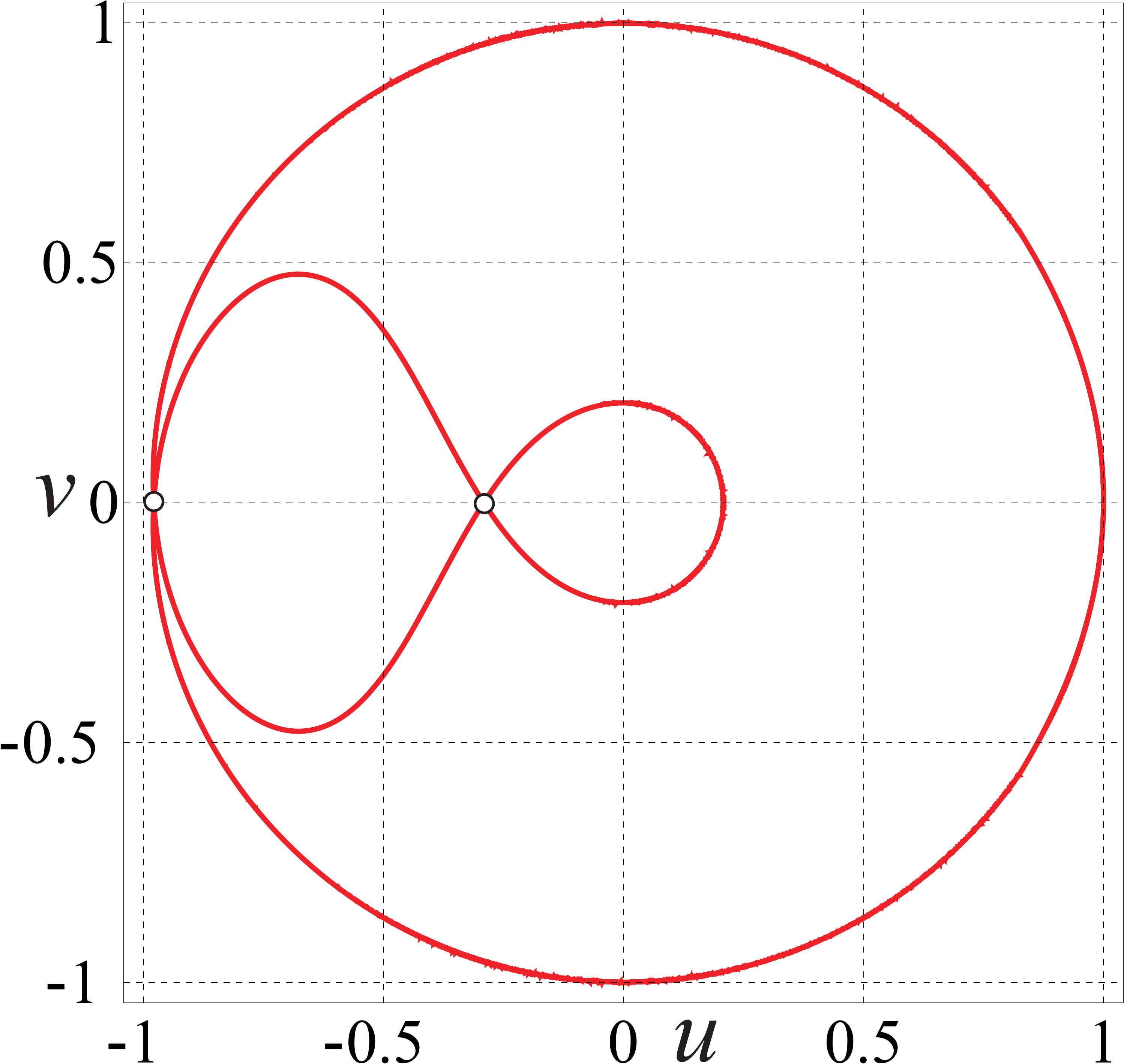}
\caption{Топологическая кривая $\mathbb S^1\,\dot{\cup}\,\mathbb S^1\,\dot{\cup}\,\mathbb S^1$.}
\label{fig3}
\end{figure}

При переходе через участок кривой $(AB)$ бифуркационной диаграммы $\Sigma_{1,b}$ при $b>3$ (см.~рис.~1) реализуется бифуркация $3\mathbb T^2 \to \mathbb S^1\times\left(\mathbb S^1\,\dot{\cup}\,\mathbb S^1\,\dot{\cup}\,\mathbb S^1\right)\to \mathbb T^2$. С помощью линий уровней приведенного гамильтониана на рис.~4 наглядно продемонстрирована указанная бифуркация трех торов в один. Если рассмотреть интегрируемое возмущение по параметру отношения интенсивностей $a$, то указанная критическая интегральная поверхность оказалась неустойчивой и распалась на два несвязных критических интегральных многообразия $S^1\times\left(\mathbb S^1\,\dot{\cup}\,\mathbb S^1\right) \cup\mathbb T^2$ и $S^1\times\left(\mathbb S^1\,\dot{\cup}\,\mathbb S^1\right)$. Этот факт также подтверждается работой \cite{oshtuzh2018}. Для наглядности на рис.~5  приведены абсолютные движения вихрей на плоскости и обмотка тора в трехмерном пространстве $\mathbb R^3(x_1,y_1,x_2)$ для следующих начальных данных $u_0 = -.37872071909766374557$; $v_0 = 0.0$; $\alpha_0 = 0$; $h=-9,f=1,03$.

\begin{figure}[!ht]
\centering
\includegraphics[width=0.7\textwidth]{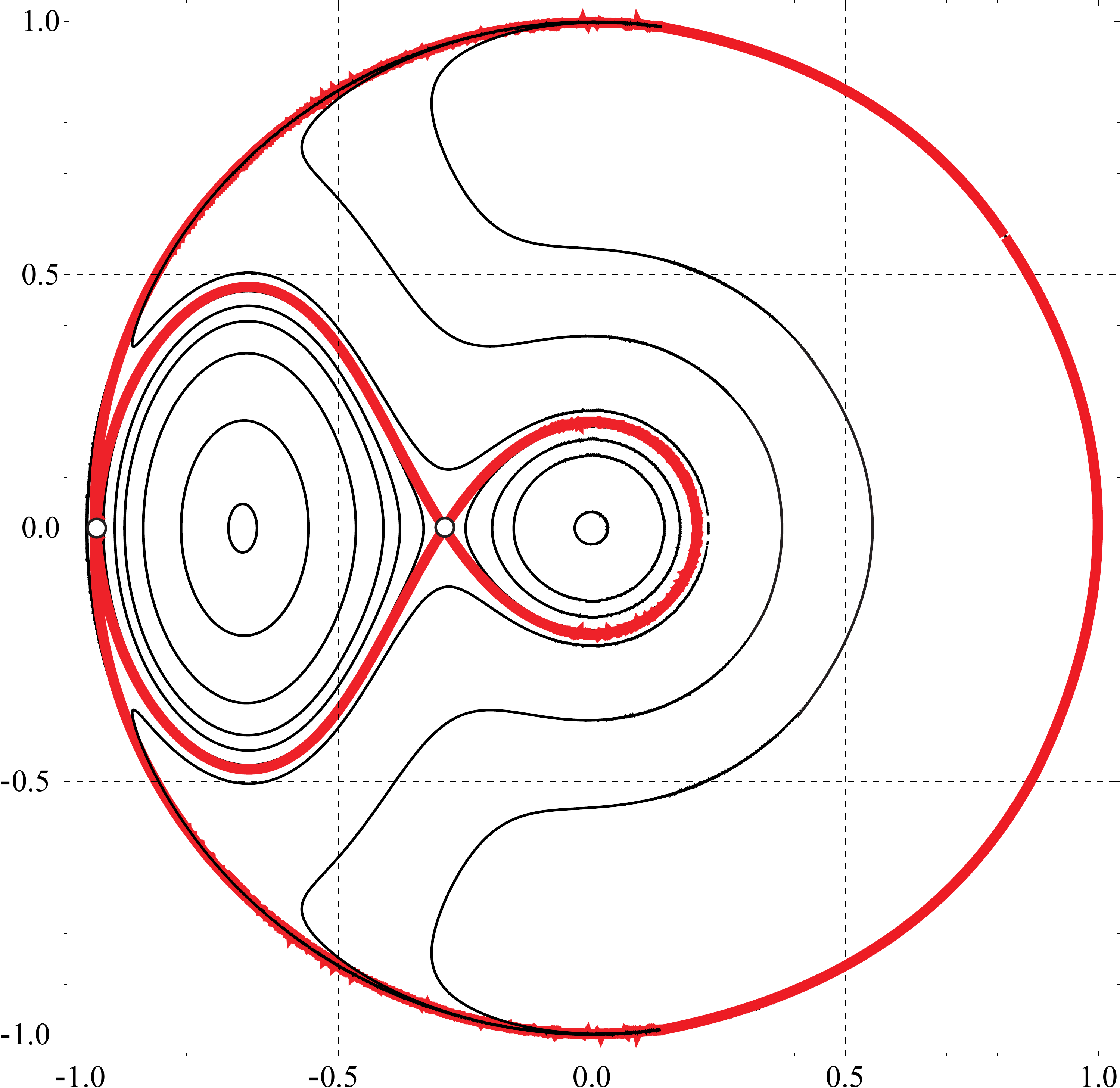}
\caption{Линии уровня приведенного гамильтониана $H_1$ для $a=1$ и $b>3$.}
\label{fig4}
\end{figure}

\begin{figure}[!ht]
\centering
\includegraphics[width=0.9\textwidth]{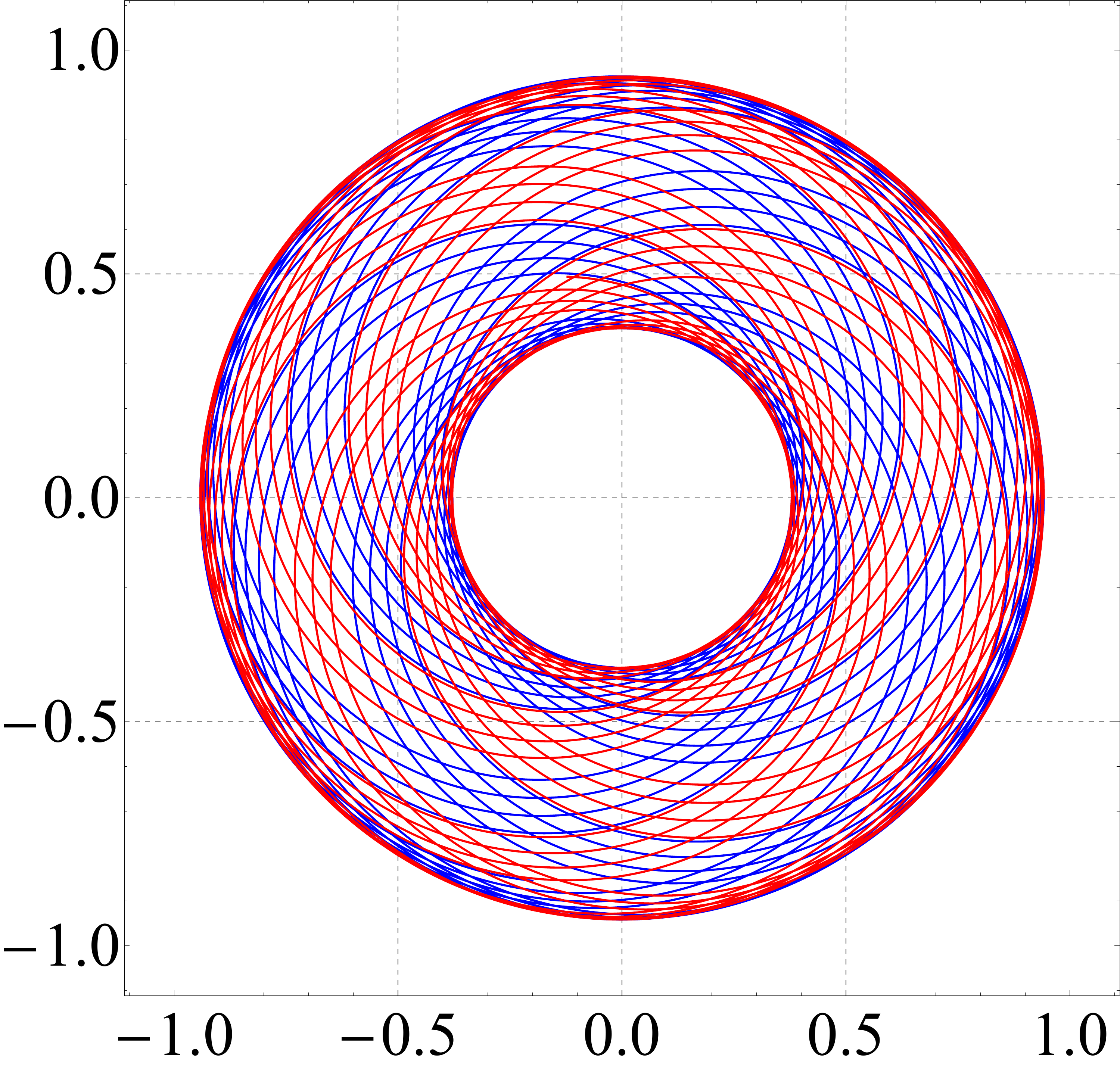}
\includegraphics[width=0.9\textwidth]{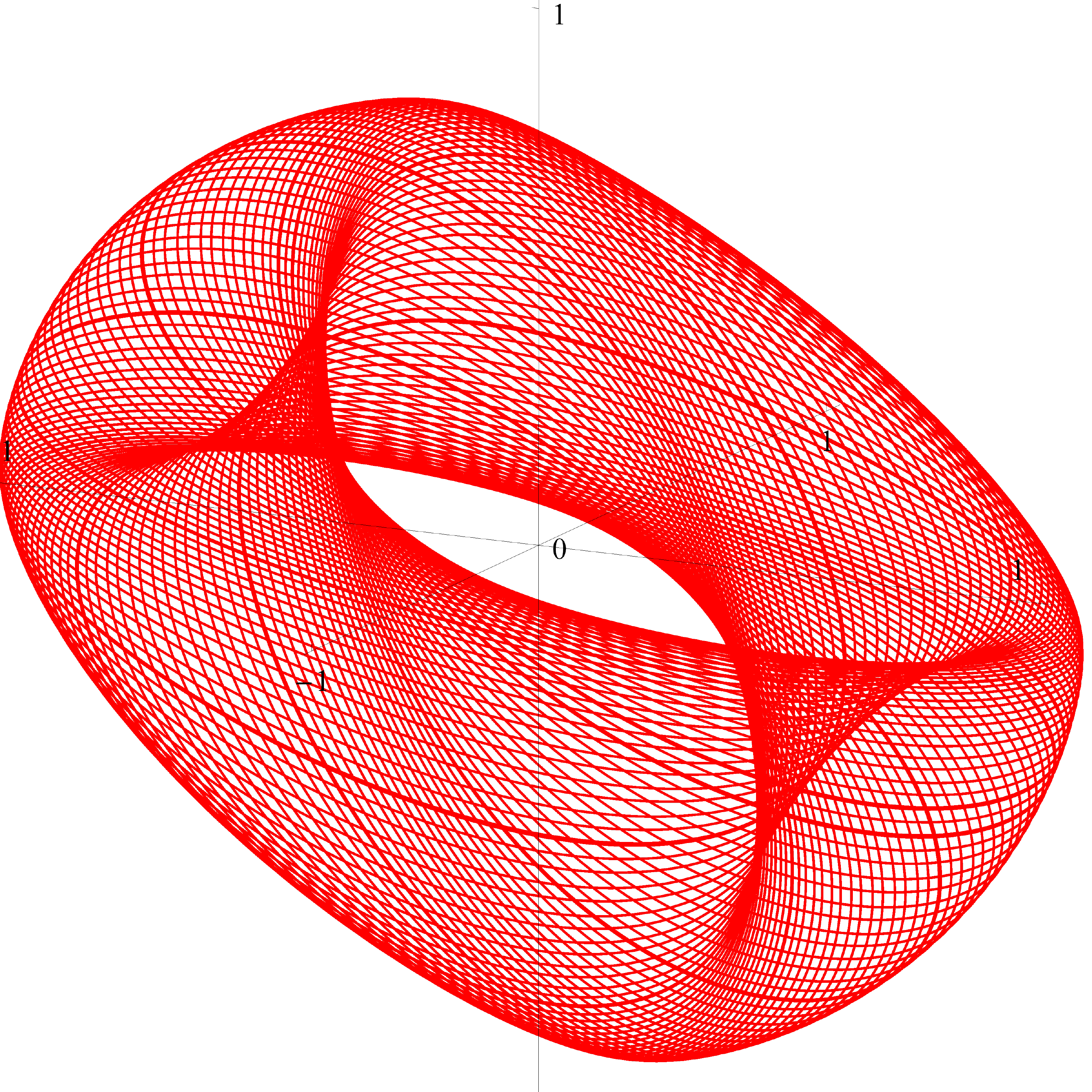}
\caption{Абсолютные движения вихрей на плоскости и в виде обмотки тора в пространстве $\mathbb R^3(x_1,y_1,x_2)$.}
\label{fig5}
\end{figure}

После разбиения фазового пространства на области, в которых количество торов остается неизменным, явного определения бифуркационной диаграммы  и самих бифуркаций можно сформулировать задачу классификации абсолютных движений вихрей, а также определения топологического типа трехмерных изоэнергетических многообразий.

\section{Благодарности}
Автор выражает благодарность А.\,В.\,Бо\-ри\-со\-ву за  плодотворные обсуждения и ценные советы, касающиеся содержания работы.
Работа выполнена при поддержке грантов РФФИ № 16-01-00170 и  17-01-00846.

\end{document}